\documentclass[12pt]{article}
\usepackage{a4}
\usepackage{amsfonts}
\usepackage{amssymb}
\usepackage{lscape}
\usepackage{cite}
\usepackage{epsfig}
\usepackage{multirow}
\usepackage[all]{xy}
\usepackage{amsmath}
\author{}

\newcommand{\drawsquare}[2]{\hbox{%
\rule{#2pt}{#1pt}\hskip-#2pt
\rule{#1pt}{#2pt}\hskip-#1pt
\rule[#1pt]{#1pt}{#2pt}}\rule[#1pt]{#2pt}{#2pt}\hskip-#2pt
\rule{#2pt}{#1pt}}
\newcommand{\Ysymm}{\raisebox{-.5pt}{\drawsquare{6.5}{0.4}}\hskip-0.4pt%
         \raisebox{-.5pt}{\drawsquare{6.5}{0.4}}}
\newcommand{\Yasymm}{\raisebox{-3.5pt}{\drawsquare{6.5}{0.4}}\hskip-6.9pt%
        \raisebox{3pt}{\drawsquare{6.5}{0.4}}}

\newcommand{\be}{\begin{equation}}
\newcommand{\ee}{\end{equation}}
\newcommand{\ba}{\begin{array}}
\newcommand{\ea}{\end{array}}
\newcommand{\bea}{\begin{eqnarray}}
\newcommand{\eea}{\end{eqnarray}}

\newcommand{\ov}{\overline}
\def\IR{\relax{\rm I\kern-.18em R}}

\def\IP{\relax{\rm I\kern-.18em P}}
\def\inbar{\vrule height1.5ex width.4pt depth0pt}
\def\IC{\relax\,\hbox{$\inbar\kern-.3em{\rm C}$}}

\def\K3{{\bf K3}}

\def\ov{\overline}

\def\n2d{\cN_{V^*}^{\otimes 2}}

\def\IC{\mathbb{C}}

\def\IR{\mathbb{R}}

\def\IP{\mathbb{P}}

\def\cN{{\mathcal N}}

\begin{document}
\title{
\begin{flushright} \vspace{-2cm}
{\small UPR-1215-T\\
\small ROM2F/2010/01\\
\vspace{-0.35cm}
} \end{flushright} \vspace{1.0cm}
The Weinberg Operator and a Lower String Scale in Orientifold Compactifications}
 \vspace{1.0cm}
\author{\small  Mirjam Cveti{\v c}$^1$, James Halverson$^1$, Paul Langacker$^2$, and  Robert Richter$^3$}

\date{}

\maketitle

\begin{center}
\emph{${}^1$Department of Physics and Astronomy, University of Pennsylvania, \\
     Philadelphia, PA 19104-6396, USA }\\\vspace{0.3cm}
 \emph{${}^2$School of Natural Science, Institute for Advanced Study,\\
 Einstein Drive, Princeton, NJ 08540}\\\vspace{0.3cm}
\emph{${}^3$ Dipartimento di Fisica and Sezione I.N.F.N. \\ Universit\`a di Roma ``Tor Vergata''\\
Via della Ricerca Scientifica, 00133 Roma, Italy}
\vspace{0.5cm}

\tt{cvetic@cvetic.hep.upenn.edu, jhal@physics.upenn.edu, pgl@sns.ias.edu, rrichter@roma2.infn.it}
\vspace{1.0cm}
\end{center}

\begin{abstract}
We investigate the interplay between the string scale and phenomenological scales in orientifold compactifications. Specifically, we discuss in generality the tension that often arises in accounting for neutrino masses, Yukawa couplings, and a $\mu$-term of the correct order and show that it often constrains the string scale $M_s$. The discussion focuses on two scenarios where, (1) the observed order of the neutrino masses are accounted for by a D-instanton induced ``stringy" Weinberg operator, or (2) effectively via the type I seesaw mechanism with an instanton induced Majorana mass term. In both scenarios, the string scale might be further constrained if the suppression factor of a single D-instanton must account for two of the phenomenological scales. For the sake of concreteness, we present phenomenologically viable quivers which exhibit these effects and perform a systematic analysis of four-stack and five-stack quivers which give rise to the exact MSSM spectrum and account for the order of the neutrino masses via the stringy Weinberg operator.
 \end{abstract}

\thispagestyle{empty} \clearpage

\section{Introduction}
One of the most challenging problems in string theory is the quest for string vacua which give rise to the phenomenology observed in nature. In addition to realizing the experimentally observed gauge symmetries and matter content, realistic string models should also account for finer details, such as the mixing angles and disparate mass scales exhibited by the Yukawa couplings. D-brane compactifications provide a promising framework for addressing such questions\cite{Blumenhagen:2005mu, Marchesano:2007de,Blumenhagen:2006ci}. In these compactifications, gauge groups appear on stacks of D-branes which fill out four-dimensional spacetime, and chiral matter appears at the intersection of two stacks.

This localization of gauge theory exhibited in D-brane compactifications, which is not present in heterotic compactifications, allows for gauge dynamics to be addressed independently of gravitational considerations. That is, many quantities relevant to particle physics, such as the chiral spectrum or the superpotential, are independent of global aspects of the geometry. Thus, for a large class of phenomenological questions it is sufficient to consider local D-brane setups which mimick a gauge theory with a desired matter field content, instead of considering a global D-brane compactification. These local setups are called D-brane quivers and this ``bottom-up" approach to model building was initiated in \cite{Aldazabal:2000sa,Antoniadis:2001np,Antoniadis:2000ena}.

Recently, there have been extensive efforts \cite{Ibanez:2008my,Leontaris:2009ci,Anastasopoulos:2009mr,Cvetic:2009yh,Cvetic:2009ez,Kiritsis:2009sf,Cvetic:2009ng,Anastasopoulos:2010} to construct semi-realistic bottom-up MSSM quivers. In theses quivers, desired Yukawa couplings are often perturbatively forbidden, since they violate global $U(1)$ selection rules which are remnants of the generalized Green-Schwarz mechanism. It has been shown \cite{Blumenhagen:2006xt,Haack:2006cy,Ibanez:2006da,Florea:2006si} that D-instantons can break these global $U(1)$ selection rules and induce perturbatively forbidden couplings (for a review, see \cite{Blumenhagen:2009qh}). In \cite{Cvetic:2009yh} (see also \cite{Cvetic:2009ng}), the authors performed a systematic bottom-up  search of D-brane quivers that exhibit the exact MSSM spectrum including three right-handed neutrinos. They investigated which quivers allow for the generation of the MSSM superpotential, perturbatively or non-perturbatively, without inducing any undesired phenomenological drawbacks, such as R-parity violating couplings or a $\mu$-term which is too large.  

One phenomenological requirement often imposed on bottom-up quivers is the existence of a mechanism which accounts for the smallness of the neutrino masses. There have been many studies of such mechanisms, both from the field theory and string points of view (for a review, see~\cite{Mohapatra:2005wg}). Many models involve small Majorana masses associated with the higher-dimensional Weinberg operator~\cite{Weinberg:1980bf} $C L H_u L H_u/M$, where $M/C \sim 10^{14}$ GeV. The Weinberg operator may be generated by integrating out heavy states in the effective four-dimensional field theory, such as a heavy right-handed Majorana neutrino (the type I seesaw mechanism). However, in D-brane compactifications a
large Majorana mass for the right-handed neutrino is likely to be due to non-perturbative D-instanton effects  \cite{Blumenhagen:2006xt,Ibanez:2006da,Cvetic:2007ku,Ibanez:2007rs,Antusch:2007jd,Cvetic:2007qj}, while in 
other string constructions it may be due to higher-dimensional operators emerging from the string compactification 
(see, e.g.,~\cite{Giedt:2005vx,Lebedev:2007hv}). It is therefore worthwhile to consider the alternative possibility that the Weinberg 
operator is generated directly by D-instantons or other stringy effects without introducing the intermediate step of a 
stringy right-handed neutrino mass\footnote{We note the analogy to the work of Klebanov and Witten \cite{Klebanov:2003my}, and the subsequent work of \cite{Cvetic:2006iz}, which calculated the superpotential contributions of stringy dimension six proton decay operators in GUT models and analyzed them relative to their effective field theoretic counterparts. Here, we consider the possibility that the Weinberg operator is a stringy effect rather than a field theoretic operator generated effectively by the Majorana mass term $N_RN_R$ or dimension four $R$-parity violating couplings, such as $q_L L d_R$.}. (Other types of string effects may lead to other possibilities, such as 
small Dirac neutrino masses \cite{Cleaver:1997nj,Cvetic:2008hi}). 

In D-brane compactifications, the scales of important superpotential couplings are often dependent on some combination of the string scale $M_s$ and the suppression factors of instantons which might induce them. Thus, it is interesting to examine whether or not a string scale different from the four-dimensional Planck scale assists in accounting for the phenomenological scales observed in nature. For example, in \cite{Anastasopoulos:2009mr} it was argued that a lower string scale may avoid the presence of a large $\mu$-term and thus may relax some of the bottom-up constraints imposed in the systematic analyses of \cite{Cvetic:2009yh,Cvetic:2009ng}. In addition to the possibility of being helpful in model building, a lower string scale is intriguing because it might give rise to stringy signatures arising from exotic matter and Regge excitations 
observable at the LHC (see, e.g.,\cite{Anastasopoulos:2006cz,Anastasopoulos:2008jt,Lust:2008qc,
Anchordoqui:2008di,Armillis:2008vp,Fucito:2008ai,Anchordoqui:2009mm,Lust:2009pz,Anchordoqui:2009ja,Burikham:2004su}
and references therein).

In this work we explore various implications of a lower string scale $M_s$ for bottom-up D-brane model building. Specifically, we analyze what a lower string scale implies for dimensionful superpotential terms, namely the $\mu$-term and the Weinberg operator, where the latter can be induced directly by a D-instanton or effectively via the type I seesaw mechanism. It has long been realized \cite{Mohapatra:2005wg,Conlon:2007zza} that a generic string induced Weinberg operator with $M\sim M_s \sim 10^{18} GeV$ and $C \leq 1$ cannot account for the observed neutrino masses. However, with a lower string scale the induced Weinberg operator may be the primary source for the neutrino masses. The D-instanton case is even more problematic \cite{Ibanez:2007rs}, since $C$ is typically exponentially suppressed. Again, however, it is possible to account for the observed neutrino masses via a D-instanton induced Weinberg operator if one lowers $M_s$ \cite{Ibanez:2007rs}.

We analyze two scenarios in detail. First, we investigate the implications on the string scale $M_s$ and bottom-up model building constraints in the case where the D-instanton induced Weinberg operator is primary source for the neutrino masses. We discuss in detail the consequences for the case where the Weinberg operator inducing instanton also induces some of the desired, but perturbatively, missing superpotential couplings. Second, we investigate the implication of a lower string scale for the case where the Weinberg operator is induced effectively via the type I seesaw mechanism with a D-instanton induced Majorana mass term.  Again we analyze the constraints on the string scale and on the bottom-up model building if the Majorana mass term generating instanton also induces one of the perturbatively forbidden superpotential couplings.

Furthermore, we perform a systematic analysis of D-brane quivers similar to the one performed in  \cite{Cvetic:2009yh,Cvetic:2009ng}, where the spectrum is the exact MSSM spectrum without right-handed neutrinos and the neutrino masses are induced by a stringy Weinberg operator.  It turns out that the absence of right-handed neutrinos requires at least four stacks of D-branes. Imposing constraints inspired by experimental observation, such as the absence of R-parity violating couplings, rules out a large class of potential D-brane quivers. Allowing for an additional D-brane stack gives many more solutions, which serve as a good starting point for future model building. 

This paper is organized as follows. In chapter \ref{chap stringy Weinberg}, we discuss the implications for the string scale if a D-instanton induced Weinberg operator is the primary source for the neutrino masses. We analyze further constraints on the string scale if an instanton which induces a desired, but perturbatively forbidden, Yukawa coupling also induces one of the dimensionful superpotential couplings, the $\mu$-term or the Weinberg operator. Furthermore, we perform a systematic analysis of multi-stack D-brane quivers that exhibit the exact MSSM spectrum, where the neutrino mass is due to a D-instanton induced Weinberg operator. The details and results of this analysis are displayed in appendix \ref{appendix}. In chapter \ref{chap stringy Weinberg}, we assume the presence of right-handed neutrinos and the Weinberg operator is induced effectively via the type I seesaw mechanism. We discuss the implications for the string scale $M_s$ if the instanton that induces one of the dimensionful quantities, the Majorana mass term for the right-handed neutrinos or the $\mu$-term, also generates a perturbatively forbidden, but desired, Yukawa coupling.

\section{The Stringy Weinberg Operator
\label{chap stringy Weinberg}}
It is known that D-instantons can induce a Weinberg operator which will give contributions to the neutrino masses \cite{Ibanez:2007rs}. Much of the discussion thus far has focused on the key fact that for the usual value of the string scale, $M_s\simeq 10^{18}GeV$, such a Weinberg operator gives contributions of at most $10^{-5}eV$, which is four orders of magnitude too small\footnote{The neutrino masses are measured to be in the range $10^{-2}-1\,\,eV$, but for simplicity's sake in the following analysis, we take the mass to be $10^{-1}\,\,eV$.}. This can be seen explicitly by examining the form of a D-instanton induced Weinberg operator,
\begin{align}
\label{eqn string weinberg operator}
e^{-S^{WB}_{ins}} \,\frac{L^i\, H_u\, L^j\, H_u}{M_s}\,\,,
\end{align}
where the stated result is under the assumptions of an instanton suppression $e^{-S^{WB}_{ins}}\simeq 1$ and $\langle H_u \rangle\simeq 100\,GeV$, which is the best case scenario, since an instanton with $e^{-S^{WB}_{ins}}<1$ further suppresses the masses. Therefore, if $M_s\simeq 10^{18} \, GeV$, the D-instanton induced  Weinberg operator only gives subleading corrections to the neutrino masses, and another mechanism must account for their observed order. In what follows, we refer to this mechanism as the stringy Weinberg operator.

Lowering the string scale has been proposed as a potential solution to the hierarchy problem \cite{ArkaniHamed:1998rs,Antoniadis:1998ig}. In such a scenario, the small string scale is due to large internal dimensions which could lower the string scale down to the TeV scale, leading to the interesting signatures at the LHC mentioned in the introduction.

A lower string scale is also an interesting possibility from the point of view of the Weinberg operator, since from (\ref{eqn string weinberg operator}) we see that with a string scale
\begin{equation}
M_s \lesssim 10^{14}\,\,\,GeV,
\end{equation}
it would give contributions on the order of $10^{-1}\,eV$, and thus could be the primary mechanism for generating neutrino masses.

In this chapter, we explore the possibility that such a Weinberg operator is the primary source of the neutrino masses.
We will see that a given quiver will require $M_s$ to be in a particular range if one hopes to obtain Yukawa couplings, a stringy Weinberg operator, and a stringy $\mu$-term of the desired order. This puts stringent constraints on the relation between the string scale and instanton suppression factors. Specifically, if the stringy Weinberg operator is to be of the observed order, then the instanton suppression factor of the Weinberg operator and the string scale are related to each other via
\begin{equation}
\frac{M_s}{e^{-S^{WB}_{ins}}} = 10^{14} \, GeV\,\,.
\label{eq: relation}
\end{equation} 
Furthermore, if the $\mu$-term is generated non-perturbatively there is also a relation between the string scale $M_s$ and the $\mu$-term instanton suppression factor, given by 
\begin{align}
 M_s \,e^{-S^{\mu}_{ins}}\, = 100 \, GeV.
 \label{eq: mu-relation}
\end{align}
From \eqref{eq: relation} and \eqref{eq: mu-relation}, it is clear that there is an additional relation relating the suppression factors of the instantons which generate the Weinberg operator and the $\mu$-term, given by
\begin{align}
e^{-S^{\mu}_{ins}} \,e^{-S^{WB}_{ins}} \simeq 10^{-12}.
\label{eq relation weinberg mu}
\end{align}
These general observations must hold in order to obtain a $\mu$-term of the desired order and neutrino masses of the observed order via a stringy Weinberg operator. The interplay between the instantons in these relations and the instantons which generate the Yukawa couplings generically give three different ranges for $M_s$, which we now discuss.

\subsection{$M_s \simeq \left( 10^3 - 10^{14}\right) \,GeV$} 
As discussed above, a stringy Weinberg operator in a compactification at the usual string scale $M_s\simeq 10^{18}\,GeV$ generates neutrino masses which are too small by four orders of magnitude. From \eqref{eq: relation}, we see that this can be compensated for if
$M_s\simeq10^{14}\,GeV$,
in which case a stringy Weinberg operator could be the primary source of neutrino mass. We emphasize, however, that this numerical value is entirely based on the choice of an $O(1)$ suppression factor for the instanton which generates the Weinberg operator. Further suppression due to the instanton could forces one to further lower the string scale to account for the observed neutrino masses. The string scale as low as the TeV scale is still compatible with experimental observations. Thus the range for the string mass is given by
\begin{align}
M_s \simeq  \left( 10^3 - 10^{14}\right)\,\,GeV\,.
\end{align}
In this range, a stringy $\mu$-term of the correct order can be generated by an instanton with suppression $e^{-S^{\mu}_{ins}}\simeq 10^{-12}-10^{-1}$.
We emphasize that, though this is the widest range of string scales compatible with a stringy Weinberg operator being the origin of the observed order of neutrino masses, a particular quiver might contain effects which constrain the range. We now discuss two such effects.

\subsection{$M_s \simeq \left(10^{9}-10^{14}\right)\,\, GeV$\label{sec middle scale}}

Often times an instanton which is required to generate a perturbatively forbidden Yukawa coupling also generates the Weinberg operator. This gives the relation \begin{align}e^{-S_{ins}}\,\,\, \equiv \,\,\, e^{-S^{WB}_{ins}}\,\,\,=\,\,\,e^{-S^{Yuk}_{ins}},\end{align} since the two suppressions are associated with the same instanton. In this case, the fact that the instanton might account for either the electron mass or the bottom-quark mass gives the lower and upper bound
\begin{align}
10^{-5} \leq e^{-S_{ins}}\leq 1.
\end{align}
Given that the suppression factor of the instanton is set by requiring that the Yukawa couplings are of the correct order, equation \eqref{eq: relation} requires that
\begin{align}
\label{scale yukawa weinberg}
M_s \simeq \left(10^{9}-10^{14} \right)\,GeV \, \, .
\end{align} 
Furthermore, since the string scale is bounded, this implies that a D-instanton induced $\mu$-term has a suppression factor $e^{-S^{\mu}_{ins}}$  in the range $10^{-12} \leq e^{-S^{\mu}_{ins}} \leq 10^{-7}$.

For the sake of conreteness, we present a five-stack quiver which exhibits this effect. Its matter spectrum and transformation behavior is given in Table \ref{table weinberg yukawa quiver}, where the $SU(3)_C$ and $SU(2)_L$ of the MSSM arise from a stack $a$ of three D-branes and a stack $b$ of two D-branes, respectively. The hypercharge is given by the linear combination $U(1)_Y=\frac{1}{6}\,U(1)_a+\frac{1}{2}\,U(1)_c+ \frac{1}{2}\,U(1)_d +\frac{1}{2}\,U(1)_e$, where the $U(1)$'s of $c$, $d$, and $e$ are associated with three other D-branes, making this a five-stack quiver. All other linear combinations of these $U(1)$'s become massive via the Green-Schwarz mechanism and survive as global symmetries which have to be obeyed on the perturbative level.

\begin{table}[h] \centering
\begin{tabular}{|c|c|c|c|c|}
\hline
 Sector & Matter Fields &  Transformation & Multiplicity & Hypercharge\\
\hline \hline
  $ab$                            & $q_L^1$  & $(a,\overline{b})$ & $1$& $\frac{1}{6}$ \\
 \hline
 $ab'$                            & $q_L^{2,3}$  & $(a,b)$ & $2$& $\frac{1}{6}$ \\
 \hline
 $ac$                            & $d_R$  & $(\overline{a},c)$  & $3 $ & $\frac{1}{3}$ \\
\hline
 $ac'$                            & $u_R^1$  & $(\overline{a},\overline{c})$  & $1 $ & $-\frac{2}{3}$ \\
\hline
 $ad'$                            & $u_R^2$  & $(\overline{a}, \ov d)$  & $1 $ & $-\frac{2}{3}$ \\
 \hline
 $ae'$                            & $u_R^3$  & $(\overline{a}, \ov e)$  & $1 $ & $-\frac{2}{3}$ \\
\hline
$bc'$                            & $H_d$  & $(\ov b,\ov c)$ & $1$ & $-\frac{1}{2}  $  \\
\hline
$bd$                            & $H_u$  & $(\ov b,d)$ & $1$ & $\frac{1}{2}  $  \\
\hline
$bd'$                            & $L^{1,2}$  & $(\ov b,\overline{d})$  & $2$& $-\frac{1}{2}$ \\
\hline
$be'$                            & $L^3$  & $(b,\ov e)$ & $1$ & $-\frac{1}{2}  $  \\
\hline
$ce'$                            & $E_R^{1,2}$  & $(c,e)$  & $2 $ & $1$ \\
\hline
$dd'$                            & $E_R^3$  & $\Ysymm_d$  & $1 $ & $1$ \\
\hline
\end{tabular}
\caption{\small {Spectrum for a quiver with $U(1)_Y=\frac{1}{6}\,U(1)_a+\frac{1}{2}\,U(1)_c+ \frac{1}{2}\,U(1)_d +\frac{1}{2}\,U(1)_e$.} }
\label{table weinberg yukawa quiver}
\end{table}
 Both, the Weinberg operator $L^{1,2} \,H_u\,L^3\,H_u$ as well as the Yukawa coupling $q_L^1\,H_u\,u_R^3$
are perturbatively forbidden since they have non-vanishing charges under some of the global $U(1)$'s, namely
\begin{align}
Q_a=0 \,\,\,\,\,\,\, Q_b=-2 \,\,\,\,\,\,\, Q_c=0 \,\,\,\,\,\,\, Q_d=1 \,\,\,\,\,\,\, Q_e=-1\,\,.
\end{align}
Couplings of this charge can be induced by an instanton $E2$ with intersection numbers\footnote{Positive intersection number $I_{E2a}$ corresponds to fermionic zero mode $\lambda_a$, tranforming as fundamental under the global $U(1)_a$.   We refer the reader to \cite{Cvetic:2009yh} for  further details on the  the non-perturbative  generation of desired superpotential terms in semi-realistic D-brane quivers.} 
\begin{align}
I_{E2a}=0 \,\,\,\,\,\,\, I_{E2b}=-1 \,\,\,\,\,\,\, I_{E2c}=0 \,\,\,\,\,\,\, I_{E2d}=1 \,\,\,\,\,\,\, I_{E2e}=-1.
\end{align}
If the presence of this instanton $E2$ is required to account for the observed  order of the charm-quark mass and neutrino masses 
\begin{align}
m^{13}_{\nu}=m^{23}_{\nu}=e^{-S_{E2}} \frac{L^{1,2}\, H_u\, L^{3}\, H_u}{M_s}\,\,,
\end{align}
then the string mass must satisfy
\begin{align}
\label{scale quiver 1}
M_s \simeq 10^{12} \,\,\, GeV\,\,.
\end{align} 
Here we assume that the suppression factor $e^{-S_{E2}}$ is of the order $10^{-2}$ to give the observed hierarchy between the top-quark and charm-quark mass.

This effect occurs in many quivers, but in this particular quiver there are two subtle issues which allow us to evade the further lowering of the string mass in (\ref{scale quiver 1}). First, note that the Weinberg operator $L^{1,2}\,H_u\,L^3\,H_u$ has two fields of positive $Q_d$ charge and one of negative $Q_d$ charge, so that the overall charge of the coupling is $Q_d=2-1=1$. Such a case allows for the coupling to be generated by an instanton with vector like zero modes \cite{Ibanez:2008my,Cvetic:2009yh,Cvetic:2009ez}, and here this instanton, $E2^{'}$, would exhibit the intersection pattern
\begin{align}
I_{E2^{'}a}=0 \,\,\,\,\,\,\, I_{E2^{'}b}=-1 \,\,\,\,\,\,\, I_{E2^{'}c}=0 \,\,\,\,\,\,\, I_{E2d^{'}}=1  \,\,\,\,\,\,\, I_{E2^{'}e}=-1\,\,\,\,\,\,\,I^{\mathcal{N}=2}_{E2^{'}d}=1\,\,.
\end{align}
Note that this instanton does have the same global $U(1)$ charges as the instanton $E2$ but does \emph{not} generate $q_L^1H_uu_R^3$, and thus its suppression factor is not bounded. Therefore, if present, its suppression factor can be tuned to account for neutrino masses of the observed order \emph{without} being forced to constrain the string mass, as in (\ref{scale quiver 1}). 

Additionally, we emphasize that $E2$ and $E2^{'}$ only generate certain couplings in the neutrino mass matrix. If the instantons which generate the other entries of this matrix can sufficiently account for the order of the neutrino masses, then one is not forced to lower the string mass, since the contributions of $E2$ and $E2^{'}$ would be small corrections to the observed neutrino masses. Therefore, for generic quivers, a detailed analysis of the neutrino mass matrix is required to determine whether or not the string scale must be constrained as in \eqref{scale quiver 1}.

\subsection{$M_s \simeq \left(10^{3} -10^{7}\right) \,\, GeV $\label{sec lower scale}}
Similarly, often times an instanton which is required to generate a perturbatively forbidden Yukawa coupling also generates the $\mu$-term. This gives the relation \begin{align}e^{-S_{ins}}\,\,\, \equiv\,\,\, e^{-S^{\mu}_{ins}}\,\,\,=\,\,\,e^{-S^{Yuk}_{ins}},\end{align} since the two suppressions are associated with the same instanton. In this case, the fact that the instanton might account for either the electron mass or the bottom-quark mass gives the lower and upper bound
\begin{align}
10^{-5} \leq e^{-S_{ins}}\leq 1.
\end{align}
Given that the suppression factor of the instanton is set by requiring that the Yukawa couplings are the correct order, equation \eqref{eq: mu-relation} requires that 
\begin{align}
\label{scale yukawa mu}
M_s \simeq \left(10^{3}-10^{7}\right) \,\,\, GeV.
\end{align} 
Furthermore via relation  \eqref{eq relation weinberg mu} the suppression factor of the D-instanton inducing the Weinberg operator has a suppression factor $e^{-S^{WB}_{ins}}$  in the range $10^{-11} \leq e^{-S^{WB}_{ins}} \leq 10^{-7}$.

\begin{table}[h] \centering
\begin{tabular}{|c|c|c|c|c|}
\hline
 Sector & Matter Fields &  Transformation & Multiplicity & Hypercharge\\
\hline \hline
  $ab$                            & $q_L^{1}$  & $(a,\overline{b})$ & $1$& $\frac{1}{6}$ \\
 \hline
 $ab'$                            & $q_L^{2,3}$  & $(a,b)$ & $2$& $\frac{1}{6}$ \\
  \hline
 $ac$                            & $d_R$  & $(\overline{a},c)$  & $3 $ & $\frac{1}{3}$ \\
\hline
 $ac'$                            & $u_R^{1,2}$  & $(\overline{a},\overline{c})$  & $2 $ & $-\frac{2}{3}$ \\
 \hline
 $ae'$                            & $u_R^{3}$  & $(\overline{a}, \ov e)$  & $1$ & $-\frac{2}{3}$ \\
\hline
$bc$                            & $H_u$  & $(\ov b,c)$ & $1$ & $\frac{1}{2}  $  \\
\hline
$bc'$                            & $H_d$  & $(\ov b,\ov c)$ & $1$ & $-\frac{1}{2}  $  \\
\hline
$bd'$                            & $L^{1,2}$  & $(\ov b,\overline{d})$  & $2$& $-\frac{1}{2}$ \\
\hline
$be$                            & $L^{3}$  & $(b,\overline{e})$  & $1$& $-\frac{1}{2}$ \\
\hline
$cd'$                            & $E_R^1$  & $(c,d)$  & $1$ & $1$ \\
\hline
$ce'$                            & $E_R^{2,3}$  & $(c,e)$  & $2$ & $1$ \\
\hline
\end{tabular}
\caption{\small {Spectrum for a quiver with $U(1)_Y=\frac{1}{6}\,U(1)_a+\frac{1}{2}\,U(1)_c+ \frac{1}{2}\,U(1)_d +\frac{1}{2}\,U(1)_e$.} }
\label{mu yukawa quiver}
\end{table}

For the sake of concreteness let us discuss a concrete quiver which realizes such a scenario. In Table \ref{mu yukawa quiver} we display the spectrum and its origin for a five-stack quiver where, as in the previous example, the hypercharge is given by the linear combination 
\begin{align}
U(1)_Y=\frac{1}{6}\,U(1)_a+\frac{1}{2}\,U(1)_c+ \frac{1}{2}\,U(1)_d +\frac{1}{2}\,U(1)_e\,\,.
\end{align}
Note that the $\mu$-term is perturbatively absent since it carries non-vanishing charge under the global $U(1)$'s
\begin{align}
Q_a=0 \,\,\,\,\,\,\, Q_b=-2 \,\,\,\,\,\,\, Q_c=0 \,\,\,\,\,\,\, Q_d=0 \,\,\,\,\,\,\, Q_e=0.
\end{align}
An instanton $E2$ with an intersection pattern 
\begin{align}
I_{E2a}=0 \,\,\,\,\,\,\, I_{E2b}=-1 \,\,\,\,\,\,\, I_{E2c}=0 \,\,\,\,\,\,\, I_{E2d}=0 \,\,\,\,\,\,\, I_{E2e}=0
\end{align}
induces the missing $\mu$-term. However, the very same instanton also generates the perturbatively missing couplings $q_L^{1}\,H_u\,u_R^{1,2}$  and $q_L^{1}\,H_d\,d_R^{1,2,3}$ which are necessary to give masses to the lightest quark family. To account for the observed mass hierarchy between the heaviest and lightest family the suppression $e^{-S_{E2}}$ factor is expected to be of the order  $10^{-5}$. This implies that the string scale is of the order 
\begin{align}
\label{scale quiver 2}
M_s \simeq 10^{7} \,\,\, GeV\,\,.
\end{align} 
Via equation \eqref{eq relation weinberg mu}, this implies that the suppression factor of the instanton which induces the Weinberg operator is expected to be of the order $e^{-S^{WB}_{ins}} \simeq 10^{-7}$.

\subsection{A Systematic Analysis of MSSM quivers}
In this chapter we have seen that the string scale must satisfy $M_s \lesssim 10^{14}\,GeV$ if one hopes to account for the observed order of neutrino masses via a stringy Weinberg operator. Furthermore, we have shown that often times the suppression factor of the instanton which induces the Weinberg operator or $\mu$-term is constrained by the requirement that the same instanton generates a Yukawa coupling of the correct order. This further constrains the value of the string mass.

In Appendix \ref{appendix}, we perform  a systematic analysis of all four-stack and five-stack quivers that exhibit the exact MSSM spectrum \emph{without} right-handed neutrinos, which is similar to the analysis performed in \cite{Cvetic:2009yh}. We impose top-down constraints which arise from global consistency conditions, such as tadpole cancellation, and bottom-up constraints which are motivated by experimental observations. The latter include, among other things, the absence of R-parity violating couplings and the absence of dimension $5$ operators that lead to rapid proton decay. The small neutrino masses are due to a stringy Weinberg operator as discussed above. Thus, for all these quivers the string mass must be $M_s \lesssim 10^{14}\,GeV$ to account for the observed neutrino masses. All four- and five-stack quivers which pass the top-down and bottom-up constraints are listed in the tables in Appendix \ref{appendix}.  We mark setups in which the string scale is further constrained due to scenarios discussed in sections \ref{sec middle scale} and \ref{sec lower scale}. The class of setups which pass all the top-down and bottom-up constraints serve as a good staring point for future D-brane model building.

\section{The Effective Weinberg Operator
\label{chap effective Weinberg}}
In the previous chapter we investigated the generation of neutrino masses via a Weinberg operator induced by a D-instanton. We saw that in order to get realistic neutrino masses the string scale has to be $M_s \lesssim 10^{14}\, GeV$ and have shown that the suppression factor of the Weinberg operator inducing instanton and the $\mu$-term inducing instanton  are related to each other via equation \eqref{eq relation weinberg mu}. Moreover, if one of the MSSM Yukawa couplings is induced by an instanton which also generates the $\mu$-term or the Weinberg operator, one obtains serious constraints on the string scale $M_s$. 

In this chapter we analyze the situation where the Weinberg operator is induced effectively via the type I seesaw mechanism. We will see that generically one can obtain realistic neutrino masses even without lowering the string scale \cite{Blumenhagen:2006xt,Ibanez:2006da,Cvetic:2007ku,Ibanez:2007rs,Antusch:2007jd,Cvetic:2007qj}. This is due to the fact that the suppression factor of the instanton which induces the Majorana masses is generically independent of the scale of the Dirac mass, whether perturbatively present or non-perturbatively generated. 

However, as encountered in \cite{Cvetic:2009yh}, the Majorana mass and Dirac mass scale are related in the case that the same instanton generates both, since the instanton suppression governs those scales. In that case the induced neutrino masses are too small to account for the observed values unless the string scale is lowered. Furthermore, we will analyze the case where one is required to lower the string scale in order to obtain a realistic $\mu$-term, analogously to section \ref{sec lower scale}.

We emphasize that the type I seesaw mechanism is the only mechanism we consider for the effective generation of the Weinberg operator, due to its relative simplicity. Other mechanisms do exist, of course, including the type II and type III seesaw mechanisms, as well as the possibility of generating the Weinberg operator via dimension four $R$-parity violating couplings, such as $q_LLd_R$ \cite{Barbier:2004ez,Chemtob:2004xr}. The latter might be seen as rather natural in the context of orientifold compactifications, where an instanton whose presence is required to generate a forbidden Yukawa coupling often generates a dimension four $R$-parity violating coupling, as well. However, the $R$-parity violating couplings generated in this way are often not suppressed enough to satisfy stringent experimental bounds, since their scale is tied to that of a Yukawa coupling via the instanton suppression factor. For these reasons, we focus on the type I seesaw mechanism, which we now discuss.

\subsection{The Generic Seesaw Mechanism
\label{sec generic seesaw}}
In the presence of a Dirac mass term which does not account for the small neutrino masses\footnote{However, see \cite{Cvetic:2008hi}  for an intriguing mechanism to obtain  small Dirac neutrino masses via D-instanton effects.}, a large Majorana mass term for the right-handed neutrinos can explain the observed small masses via the seesaw mechanism.   
When generated by D-instantons, these mass terms take the form
\begin{align}
e^{-S^{Dirac}_{ins}}L^I \,H_u \,N_R^J\qquad \qquad  e^{-S^{Majo}_{ins}}\,\,M_s\,N_R^I \,N_R^J\,\,.
\end{align}
These terms give a mass matrix of the form\footnote{For simplicity we display the neutrino mass matrix for one family only.}
\begin{align}
m_{\nu}= \left( \begin{array}{cc} 0 & e^{-S^{Dirac}_{ins}} \langle H_u \rangle\\
e^{-S^{Dirac}_{ins}}\langle H_u \rangle & e^{-S^{Majo}_{ins}} M_s \\
\end{array}\right)\,\,,
\label{eq generic seesaw mass matrix}
\end{align}
where $\langle H_u \rangle$ denotes the VEV of the Higgs field and $M_s$ is the
string mass. If the Majorana mass term for the right-handed neutrinos is much larger than the Dirac mass term, then the mass eigenvalues of $m_{\nu}$ are of the order
\begin{align}
m^1_{\nu}= \,\frac{e^{-2\,S^{Dirac}_{ins}}\,{\langle H_u \rangle}^2}{e^{-S^{Majo}_{ins} }\,M_s} \qquad \text{and}
\qquad m^2_{\nu}= e^{-S^{Majo}_{ins}}\,M_s\,\,.
\end{align}
Taking $\langle H_u \rangle\simeq 100 \, GeV$ and the observed neutrino mass $m^1_{\nu} \simeq 10^{-1} \, eV $, this relates the string scale $M_s$ to the two suppression factors $e^{-\,S^{Dirac}_{ins}}$  and $e^{-S^{Majo}_{ins}}$ via the relation
\begin{align}
M_s \simeq \frac{e^{-2\,S^{Dirac}_{ins} } } {e^{-S^{Majo}_{ins} }} \, 10^{14}\, GeV\,\,.
\label{eq relation M_s suppression seesaw}
\end{align}
Note that in contrast to the stringy Weinberg operator discussed in the previous section, here we do not have to lower the string scale to obtain realistic neutrino masses.
Thus, even for the generic string scale $M_s\simeq 10^{18}\, GeV$, we obtain realistic neutrino masses if the suppression factors satisfy
\begin{align}
\label{eq condition seesaw generic string scale}
10^{4}\simeq \frac{e^{-2\,S^{Dirac}_{ins} } } {e^{-S^{Majo}_{ins} }} \,\,.
\end{align}
Of course, being able to satisfy this crucially depends on the fact that the instanton suppression factors are not related to one another.

\subsection{The Seesaw with Lower String Scale
\label{sec seesaw lower string scale}}

We saw that for a Weinberg operator induced by the type I seesaw mechanism, one can generically obtain neutrino masses in the observed range for the string scale 
$M_s\simeq 10^{18}\, GeV$, as long as the suppression factors of the Dirac and Majorana mass inducing instantons satisfy the condition \eqref{eq condition seesaw generic string scale}. This is due to the fact that  the seesaw neutrino masses contain two parameters, $e^{-S^{Dirac}_{ins}}$ and $e^{-S^{Majo}_{ins}}$,  while the neutrino masses generated via a D-instanton induced Weinberg operator depend only one parameter, namely $e^{-S^{WB}_{ins}}$.
 
However, the situation changes if the same instanton generates both the Dirac mass and Majorana mass terms.
Then we have  
\begin{align}
 e^{-S_{ins}}  \,\,\, \equiv \,\,\,   e^{-S^{Dirac}_{ins}} \,\,\,=\,\,\,e^{-S^{Majo}_{ins}} 
\end{align}
in equation \eqref{eq generic seesaw mass matrix}, and the seesaw masses take the form
\begin{align}
m^1_{\nu}= e^{-S_{ins}}\,\frac{{\langle H_u \rangle}^2}{M_s} \qquad \text{and}
\qquad m^2_{\nu}= e^{-S_{ins}}\,M_s\,\,.
\end{align}
Note that the light mass eigenvalue is in a form similar to the one encountered in the previous chapter, where the neutrino masses arose from a Weinberg operator induced by a D-instanton. With the generic values $M_s \simeq 10^{18} \,GeV$, $\langle H_u \rangle \simeq 100\, GeV$, and an $O(1)$
instanton suppression factor, this gives neutrino masses of order $10^{-5}\, eV$, which is too small by a few orders of magnitude. Thus, in this case, the type I seesaw mechanism cannot account for the observed order of the neutrino masses unless $M_s$ is significantly lower than $10^{18}\, GeV$. The relation between the string scale and the suppression factor which has to be satisfied in order to obtain neutrino masses of the order $10^{-1}\, eV$ is
\begin{align}
\frac{M_s}{e^{-S_{ins}}}\, = 10^{14} \, GeV\,\,.
\end{align}
This is precisely the same relation as \eqref{eq: relation}, except that now it has arisen as a special case of the type I seesaw mechanism, rather than from a stringy Weinberg operator. As before, we know that $e^{-S_{ins}}$ is at most $O(1)$, so that in this scenario the string scale has to satisfy $M_s \lesssim 10^{14}\,GeV$.

We now present a four stack quiver which realizes such a scenario. The hypercharge for this quiver is  $U(1)_Y=-\frac{1}{3}\,U(1)_a-\frac{1}{2}\,U(1)_b$, and its matter content and transformation behavior is given in Table \ref{lower seesaw quiver}. Note that this quiver was encountered in the systematic analysis performed in \cite{Cvetic:2009yh} and corresponds to solution 3 in Table 7 of  \cite{Cvetic:2009yh}.
\begin{table}[h] \centering
\begin{tabular}{|c|c|c|c|c|}
\hline
 Sector & Matter Fields &  Transformation & Multiplicity & Hypercharge\\
\hline \hline
  $ab$                            & $q_L$  & $(a,\overline{b})$ & $3$& $\frac{1}{6}$ \\
 \hline
 $ad'$                            & $d_R$  & $(\overline{a},\ov d)$  & $3 $ & $\frac{1}{3}$ \\
 \hline
 $aa'$                            & $u_R$  & $\Yasymm_a$  & $3 $ & $-\frac{2}{3}$ \\
\hline
$bc'$                            & $H_u \,\,\, H_d$  & $(\ov b, \ov c) \,\,\, (b,c)$ & $1$ & $\frac{1}{2} \,\,\,-\frac{1}{2}  $  \\
\hline
$bd$                            & $L$  & $(b,\overline{d})$  & $3$& $-\frac{1}{2}$ \\
\hline
$bb'$                            & $E_R$  & ${\ov \Yasymm}_b$  & $3$ & $1$ \\
\hline
$cd'$                            & $N_R$  & $(\ov c, \ov d)$ & $3$ & $0$ \\
\hline
\end{tabular}
\caption{\small {Spectrum for a quiver with $U(1)_Y=-\frac{1}{3}\,U(1)_a-\frac{1}{2}\,U(1)_b.$} }
\label{lower seesaw quiver}
\end{table}
In this quiver, the Dirac mass term $L^I\, H_u\, N_R^J$ and Majorana mass term $N_R^I\,N_R^J$ both have charge
\begin{align}
Q_a=0 \,\,\,\,\,\,\, Q_b=0 \,\,\,\,\,\,\, Q_c=-2 \,\,\,\,\,\,\, Q_d=-2
\end{align}
under the global $U(1)$'s. Couplings of this charge can be induced by an instanton $E2$ with intersection pattern
\begin{align}
I_{E2a}=0 \,\,\,\,\,\,\, I_{E2b}=0 \,\,\,\,\,\,\, I_{E2c}=-2 \,\,\,\,\,\,\, I_{E2d}=-2.
\end{align}
This is precisely the case discussed above, where the same instanton generates both the Dirac mass term and the Majorana mass term. Therefore there is an upper bound $M_s \lesssim 10^{14}\,GeV$ in order to obtain neutrino masses of the correct order. Since the $\mu$-term is perturbatively present and $E2$ does not generate any of the Yukawa couplings, there is no interplay between these effects which would further constrain $M_s$.

\subsection{Lower string scale due to $\mu$-term
\label{sec seesaw mu-term}}
As discussed in section \ref{sec lower scale}, one might have to lower the string scale in order to obtain a realistic $\mu$-term. This happens if one of the instantons which induces a perturbatively forbidden, but desired, Yukawa coupling also generates the $\mu$-term \cite{Anastasopoulos:2009mr}. In that case, 
\begin{align}
e^{-S_{ins}}\equiv e^{-S^{\mu}_{ins}} \simeq e^{-S^{Yuk}_{ins}}
\end{align}
and in order to get realistic mass hierarchies we have
\begin{align}
10^{-5} \leq e^{-S_{ins}}\leq 1\,\,.
\end{align}
To obtain a $\mu$-term of the correct order, this requires
\begin{align}
M_s \simeq \left(10^{3} - 10^{7} \right)\, GeV\,\,.
\label{mu ms range}
\end{align}
From equation \eqref{eq relation M_s suppression seesaw}, we see that a string mass in this range requires the suppression factors $e^{-S^{Dirac}_{ins}}$ and $e^{-S^{Majo}_{ins}}$ to satisfy
\begin{align}
10^{-11} \leq  \frac{e^{-2\,S^{Dirac}_{ins} } } {e^{-S^{Majo}_{ins} }} \leq 10^{-7}\,\,.
\label{eq relation seesaw mu-term }
\end{align}
Note that if the Dirac neutrino mass term is realized perturbatively, then $e^{-S^{Dirac}_{ins}} \simeq 1$ and the Majorana mass term is not large enough to account for the small neutrino masses via the seesaw mechanism. In this case they would be larger than observed. Furthermore, if the Dirac neutrino mass term is induced by an instanton which also generates another perturbatively forbidden Yukawa coupling, then $e^{-S^{Dirac}_{ins}} = e^{-S^{Yuk}_{ins}}$, and $e^{-S^{Majo}_{ins} }$ can only live in the small window
\begin{equation}
10^{-3}\lesssim \, e^{-S^{Majo}_{ins} } \, \lesssim 1.
\end{equation}
This is a consequence of the relation between the Dirac mass term and Yukawa coupling inducing instanton suppressions, the relation \eqref{eq relation seesaw mu-term }, and the fact that any instanton suppression is at most $O(1)$.
If, on the other hand, the instanton which generates the Dirac mass term does not generate a perturbatively forbidden Yukawa coupling or the Majorana mass term, then its suppression factor is unconstrained and one can obtain realistic neutrino masses, in accord with \eqref{eq relation seesaw mu-term }.

As an example, we present a four-stack quiver with the hypercharge embedding $U(1)_Y=\frac{1}{6}\,U(1)_a+\frac{1}{2}\,U(1)_c+\frac{1}{2}\,U(1)_d$, whose matter content and transformation behavior is given in Table \ref{mu seesaw quiver}.
\begin{table}[h] \centering
\begin{tabular}{|c|c|c|c|c|}
\hline
 Sector & Matter Fields &  Transformation & Multiplicity & Hypercharge\\
\hline \hline
  $ab$                            & $q_L$  & $(a,\overline{b})$ & $3$& $\frac{1}{6}$ \\
 \hline
 $ac$                            & $d_R$  & $(\overline{a}, c)$  & $3 $ & $\frac{1}{3}$ \\
 \hline
 $ac'$                            & $u_R^1$  & $(\overline{a},\ov c)$  & $1 $ & $-\frac{2}{3}$ \\
 \hline
 $ad'$                            & $u_R^{2,3}$  & $(\overline{a},\ov d)$  & $2 $ & $-\frac{2}{3}$ \\
\hline
$bc'$                            & $H_d$  & $ (\ov b,\ov c)$ & $1$ & $-\frac{1}{2}  $  \\
\hline
$bd$                            & $L$  & $(b,\overline{d})$  & $3$& $-\frac{1}{2}$ \\
\hline
$bd'$                            & $H_u$  & $ (b,d)$ & $1$ & $\frac{1}{2}  $  \\
\hline
$bb'$                            & $N_R$  & ${\ov \Yasymm}_b$  & $3$ & $0$ \\
\hline
$cc'$                            & $E_R^2$  & $\Ysymm_c$ & $1$ & $1$ \\
\hline
$dd'$                            & $E_R^{2,3}$  & $\Ysymm_d$ & $2$ & $1$ \\
\hline
\end{tabular}
\caption{\small {Spectrum for a quiver with $U(1)_Y=\frac{1}{6}\,U(1)_a+\frac{1}{2}\,U(1)_c+\frac{1}{2}\,U(1)_d.$} }
\label{mu seesaw quiver}
\end{table}
In this quiver, the $\mu$-term $H_u\, H_d$ and the Yukawa couplings $q^I_L\, H_u\,  u^1_R$ have global $U(1)$ charge 
\begin{align}
Q_a=0 \,\,\,\,\,\,\, Q_b=0 \,\,\,\,\,\,\, Q_c=-1 \,\,\,\,\,\,\, Q_d=1\,\,.
\end{align}
 Couplings of this charge can be induced by an instanton $E2_1$ with intersection pattern
\begin{align}
I_{E2a}=0 \,\,\,\,\,\,\, I_{E2b}=0 \,\,\,\,\,\,\, I_{E2c}=-1 \,\,\,\,\,\,\, I_{E2d}=1.
\end{align}
If the instanton $E2$ is to account for the observed up-quark mass, then the suppression factor is fixed to be  $e^{-S_{E2}} \simeq 10^{-5}$ and the string scale must be of the order
\begin{align}
M_s \simeq  10^7 \,GeV.
\end{align}
In this quiver the Dirac mass term $L\,H_u\,N_R$ is perturbatively realized and, as discussed above, one cannot obtain realistic neutrino masses via the seesaw mechanism. It turns out that if one allows for a lower string scale and thus relaxes the conditions on the non-perturbative generation of the $\mu$-term\footnote{Here we implemented in addition to the constraints laid out in \cite{Cvetic:2009yh} also constraints on dimension five operators discussed in \cite{Kiritsis:2009sf,Cvetic:2009ez,Cvetic:2009ng}. }, there are no additional four-stack quivers compared to the ones found in \cite{Cvetic:2009yh}. It is for the reason that we have presented a quiver which does not allow for realistic neutrino masses. However we expect the situation to change if the MSSM realization is based on five D-brane stacks. In that case we expect that lowering the string scale to account for a realistic non-perturbative $\mu$-term will increase the number of realistic bottom-up MSSM-quivers. 

\section{Conclusions}

Perhaps the most astonishing experimental particle physics results in recent memory is the the existence of very small neutrino masses, which differ from the top-quark mass by over ten orders of magnitude. The existence of such a discrepancy in nature is fascinating, and much effort has been dedicated to investigating possible explanations. One answer, of course, is that the masses``are what they are." But this isn't very satisfying, particularly since another explanation might give insight into fundamental theory or beyond the standard model effective theory. Many effective and string theoretic explanations have been developed to this end, including a number of ``seesaw" mechanisms and neutrino mass terms directly induced by D-instanton effects in type II superstring theory.

The small scale of the neutrino masses is not the only interesting experimentally observed scale in particle physics, of course. The Yukawa couplings, with their disparate mass scales and mixing angles, are also fascinating and furthermore it is important to have a $\mu$-term of the correct order in the MSSM. 
As shown in \cite{Anastasopoulos:2009mr,Cvetic:2009ez}, D-instanton effects may account for these observed mass hierarchies. However, often times a D-instanton induces more than one of the perturbatively forbidden couplings, which would relate their scales and therefore might pose phenomenological problems\cite{Ibanez:2008my,Cvetic:2009yh,Cvetic:2009ez}. The string scale $M_s$ affects the scale of dimensionful parameters, though, and therefore a lower string scale might alleviate these problems \cite{Anastasopoulos:2009mr}. 

In this work we investigate the implications of a lower string scale for bottom-up D-brane model building. We show that a lower string scale allows for a D-instanton induced Weinberg operator to be the primary source for the neutrino masses. For a generic string scale of $10^{18}\, GeV$, the latter would generate neutrino masses much smaller than the observed ones \cite{Ibanez:2007rs}. Thus, a lower string scale provides an option to obtain many semi-realistic bottom-up D-brane quivers. However, a lower string scale sometimes also poses a serious problem. Specifically, if the Dirac mass term is realized perturbatively and the Majorana mass is smaller due to a lower string scale, then neutrino masses of the observed order cannot be obtained via the type I seesaw mechanism.

In chapter \ref{chap stringy Weinberg}, we discuss the scenario where the neutrino masses are due to a D-instanton induced Weinberg operator. We show that the string scale has to be lower than $10^{14}\, GeV$ to obtain realistic neutrino masses. We further investigate how the string scale might be further constrained to account for the observed mass hierarchies if one D-instanton induces multiple desired superpotential couplings. Finally we perform a systematic analysis similar to the one performed in  
\cite{Cvetic:2009yh,Cvetic:2009ez} of multi D-brane stack quivers, which exhibit the exact MSSM spectrum but without right-handed neutrinos, where the observed order of the neutrino masses are due to a stringy Weinberg operator.    

In chapter \ref{chap effective Weinberg}, we investigate the generation of an effective Weinberg operator via the type I seesaw mechanism. While generically the type I seesaw mechanism can account for the observed neutrino masses even with the usual string scale, it sometimes happens that the same instanton which induces the Majorana mass term for the right-handed neutrinos also generates the Dirac neutrino mass term\cite{Cvetic:2009yh}. In that case one encounters a situation anologous to the stringy Weinberg operator and one is forced to lower the string scale to account for the observed neutrino masses. As in chapter \ref{chap stringy Weinberg}, we further analyze the implications on the string scale if a D-instanton induces multiple desired, but perturbatively missing, superpotential couplings. We encounter a large tension between getting realistic neutrino masses and a $\mu$-term of the desired order if the latter is induced by a D-instanton which also induces one of the perturbatively missing Yukawa couplings.  
 
\vspace{1.5cm}
{\bf Acknowledgments}\\

We thank M. Ambroso, P. Anastasopoulos, M. Bianchi, F. Fucito, I. Garc\'ia-Etxebarria, G. K. Leontaris, A. Lionetto and J.F. Morales for useful discussions.
The work of M.C. and J.H. is supported by the DOE  Grant DOE-EY-76-02-3071, the NSF RTG grant DMS-0636606 and the Fay R. and Eugene L. Langberg Chair.
The work of P.L. is supported by the IBM Einstein Fellowship and by the NSF grant PHY-0503584. R.R. thanks the University of Pennsylvania for hospitality during this work.
 
\newpage
\appendix
\section{Results of a Systematic Analysis \label{appendix}}
In this appendix we present the results of a systematic analysis of four-stack and five-stack quivers that exhibit the exact MSSM spectrum \emph{without} right-handed neutrinos. Similarly to the analysis performed in \cite{Cvetic:2009yh}, we impose top-down and bottom-up constraints, which arise from global consistency conditions of D-brane compactifications and experimental observations, respectively. All quivers presented allow for the observed order of the neutrino masses to be accounted for by a stringy Weinberg operator, discussed in chapter \ref{chap stringy Weinberg}. 

Below we display all the top-down and bottom-up constraints we impose in the systematic analysis. The four- and five-stack quivers passing these constraints serve as a good starting point for future D-brane model building. The constraints are as follows:
\begin{itemize}
\item[$\bullet$] Tadpole cancellation, which is a condition on the cycles that the D-branes wrap, imposes constraints on the transformation behavior of the chiral matter. For a stack $a$ of $N_a$ D-branes  with $N_a>1$, the constraints read
\begin{align}
\#(a) - \#({\ov a})  + (N_a-4)\#( \, \Yasymm_a) + (N_a+4) \#
(\Ysymm_a)=0 \,\,,\label{eq constraint1}
\end{align}
which is equal to the anomaly cancellation for $N_a >2$ but also holds for $N_a=2$.  For $N_a=1$ the constraint is slightly modified due to the absence of anti-symmetric tensors and takes the form
\begin{align}
\#(a) - \#({\ov a}) + 5 \# (\Ysymm_a)=0  \qquad \text{mod} \,3\,\,.
\label{eq constraint2}
\end{align}
We require the constraints to be satisfied for all D-brane stacks by the chiral matter content of the respective quiver.
\item[$\bullet$] In a fashion similar to tadpole cancellation, the presence of a massless $U(1)_Y$ puts constraints on the cycles that the D-branes wrap. These again imply constraints on the transformation behaviour of the chiral matter, which are given by
\begin{align}
 \sum_{x \neq a} q_x\,
N_x \#(a,{\ov x}) - \sum_{x \neq a} q_x\, N_x \#(a,x) = q_a\,N_a \,\Big(\#(\Ysymm_a) + \# (\, \Yasymm_a)\Big) \,\,
\label{eq massless constraint non-abelian}
\end{align}
for $N_a>1$ and
\begin{align}\sum_{x \neq a} q_x\, N_x \#(a,{\ov x}) -
\sum_{x \neq a} q_x\, N_x \#(a,x)= q_a\,\frac{\#(a) - \#({\ov a}) + 8 \#
(\Ysymm_a)}{3}  
\label{eq massless constraint abelian}
\end{align}
for $N_a=1$. We require these constraints to be satisfied by the matter content of the respective quiver.
\item[$\bullet$] We require that the chiral spectrum of the quivers is the exact MSSM spectrum \emph{without} right-handed neutrinos. Thus the top-down constraints arising from tadpole cancellation and the presence of a massless $U(1)_Y$ have to be satisfied within the MSSM spectrum.
\item[$\bullet$] The MSSM superpotential couplings 
\begin{align}  
q_L\, H_u \,u_{R} \qquad  q_L\, H_d \,d_{R} \qquad L \, H_d \, E_R \qquad H_u\, H_d
\end{align}
are either realized perturbatively or in case they violate global $U(1)$ selection rules and thus are perturbatively forbidden they will be induced by D-instantons, in such a way that all three families of quarks and charged leptons acquire masses. 
\item[$\bullet$] The small neutrino masses are due to a D-instanton induced Weinberg operator as discussed in chapter \ref{chap stringy Weinberg}.

\item[$\bullet$] We require the absence of the R-parity violating couplings
\begin{equation}
d_R\, d_R\, u_R \qquad  L \, L\, E_R \qquad  q_L\, L\, d_R \qquad  L\, H_u
\end{equation}
on the perturbative level. Furthermore, we require that they are not generated by an instanton whose presence is required to generate a perturbatively forbidden, but desired, MSSM superpotential coupling.

\item[$\bullet$] We require the absence of the dangerous dimension $5$ proton decay operators
\begin{equation}
u_R\,u_R\,d_R\,E_R \qquad  \text{and} \qquad q_L\,q_L\, q_L\, L
\end{equation}
on the perturbative level. Furthermore, we require that they are not generated by an instanton whose presence is required to generate a perturbatively forbidden, but desired, MSSM superpotential coupling.

\item[$\bullet$] For a very low string scale also the dimension $5$ operator 
\begin{equation}
\frac{\kappa}{M_s}\,\,q_L\,q_L\, q_L\, H_d
\end{equation}
might lead to dangerously high rates for $\Delta B = \pm 2$ processes\footnote{The rate for proton decay induced by the combination of this operator and the Weinberg operator is negligibly small because it would require an additional source of $R$-parity violation.} such as
neutron oscillations or nucleon-nucleon annihilation in a nucleus. There is
some theoretical uncertainty in the experimental bound \cite{Barbier:2004ez}, but a conservative
estimate is 
\begin{equation}
\frac{\kappa\, \langle H_d \rangle }{M_s} \,\leq 10^{-8}\,\,,
\end{equation}
where $M_s$ denotes the string scale and $\langle H_d \rangle$ is the vev of $H_d$.
For a small string scale one has to ensure that this operator is perturbatively absent and moreover check if none of the instantons whose presence is required to induce some of the perturbatively missing but desired Yukawa couplings does not generate also this operator.
\item[$\bullet$] We require that the top-quark Yukawa coupling is realized perturbatively, or in case it is induced non-perturbatively all other Yukawa couplings are also perturbatively forbidden. In the latter case the quiver can still account for the observed hierarchy between the top-quark mass and all other matter field masses.
\item[$\bullet$] We rule out any quiver which clearly exhibits too much mixing between different quark families.

\end{itemize}

The quivers presented in this appendix represent all three-stack, four-stack, and five-stack MSSM quivers which satisfy all of these constraints.

\subsection*{Three-Stack Quivers}

For the three-stack quivers one has two potential hypercharge embeddings compatible with the top-down constraints, namely
\begin{align*}
 U(1)_Y= \frac{1}{6}\, U(1)_a + \frac{1}{2} \,U(1)_c \qquad \qquad U(1)_Y= -\frac{1}{3}\, U(1)_a - \frac{1}{2} \,U(1)_b \,\,.
\end{align*}
For both embeddings one can show that there exists no quiver  which passes all the bottom-up constraints laid out above.
\subsection*{Four-Stack Quivers}

Let us turn to the four-stack quivers. Below we  display all possible hypercharge embeddings which are in agreement with the hypercharge assignments of the MSSM matter fields \cite{Anastasopoulos:2006da}.  Requiring that each D-brane stack is populated allows for the following hypercharge embeddings 
\begin{itemize}
\item[$\bullet$] $U(1)_Y= \frac{1}{6}\, U(1)_a + \frac{1}{2} \,U(1)_c + \frac{1}{2} \,U(1)_d$
\item[$\bullet$] $U(1)_Y= \frac{1}{6}\, U(1)_a + \frac{1}{2} \,U(1)_c + \frac{3}{2} \,U(1)_d$
\item[$\bullet$] $U(1)_Y= -\frac{1}{3}\, U(1)_a - \frac{1}{2} \,U(1)_b $
\item[$\bullet$] $U(1)_Y= -\frac{1}{3}\, U(1)_a - \frac{1}{2} \,U(1)_b +\frac{1}{2} \, U(1)_d$
\item[$\bullet$] $U(1)_Y= -\frac{1}{3}\, U(1)_a - \frac{1}{2} \,U(1)_b + U(1)_d$\,\,.
\end{itemize}

Only for two embeddings do we find quivers which pass all the constraints that we impose. They are listed in Tables \ref{four stack two} and \ref{four stack three}.

\begin{table}[h]
\hspace{-.2cm}
\begin{tabular}{|c|c|c|c|c|c|c|c|c|c|c|}\hline
		\multirow{2}{*}{Solution \#}&\multicolumn{1}{|c}{$q_L$} & \multicolumn{1}{|c}{$d_R$} & \multicolumn{2}{|c}{$u_R$} & \multicolumn{1}{|c}{$L$} & \multicolumn{2}{|c}{$E_R$}
		& \multicolumn{1}{|c}{$H_u$} & \multicolumn{2}{|c|}{$H_d$}\\ \cline{2-11}
		&$(a,\ov{b})$ 
		&$(\ov{a},\ov{c})$
		&$(\ov{a},\ov{d})$&$\Yasymm_a$
		&$(b,\ov{c})$
		&$(c,d)$&$\ov{\Yasymm}_b$
		&$(b,d)$
		&$(b,c)$&$(\ov{b},\ov{d})$
		\\
		\hline\hline			
		1&3&3&3&0&3&1&2&1&1&0\\ \hline
		2&3&3&3&0&3&0&3&1&0&1\\ \hline
		3&3&3&0&3&3&0&3&1&0&1\\ \hline
\end{tabular}
\caption{Spectrum for the solutions with $U(1)_Y= -\frac{1}{3}\, U(1)_a - \frac{1}{2} \,U(1)_b + U(1)_d $.
\label{four stack two}
}
\end{table}

\begin{table}[h]
\hspace{1.5cm}
\begin{tabular}{|c|c|c|c|c|c|c|c|}\hline
		\multirow{2}{*}{Solution \#}&\multicolumn{1}{|c}{$q_L$} & \multicolumn{1}{|c}{$d_R$} & \multicolumn{1}{|c}{$u_R$} & \multicolumn{1}{|c}{$L$} & \multicolumn{1}{|c}{$E_R$}
		& \multicolumn{1}{|c}{$H_u$} & \multicolumn{1}{|c|}{$H_d$}\\ \cline{2-8}
		&$(a,\ov{b})$ 
		&$(\ov{a},\ov{d})$
		&$\Yasymm_a$
		&$(b,\ov{d})$
		&$\ov{\Yasymm}_b$
		&$(\ov{b},\ov{c})$
		&$(b,c)$
		\\
		\hline\hline			
		1&3&3&3&3&3&1&1\\ \hline
\end{tabular}
\caption{Spectrum for the solutions with $U(1)_Y= -\frac{1}{3}\, U(1)_a - \frac{1}{2} \,U(1)_b $.
\label{four stack three}
}
\end{table}




\subsection*{Five-Stack Quivers}
For the five stack quivers we have the following hypercharge embeddings, which are in agreement with the hypercharge assignment of the MSSM matter fields. Again we require that all D-brane stacks are populated, which results into the following choices  

\begin{itemize}
\item[$\bullet$] $U(1)_Y= \frac{1}{6}\, U(1)_a + \frac{1}{2} \,U(1)_c + \frac{1}{2} \,U(1)_d+\frac{1}{2} \,U(1)_e$
\item[$\bullet$] $U(1)_Y= \frac{1}{6}\, U(1)_a + \frac{1}{2} \,U(1)_c + \frac{1}{2} \,U(1)_d+\frac{3}{2} \,U(1)_e$
\item[$\bullet$] $U(1)_Y= \frac{1}{6}\, U(1)_a + \frac{1}{2} \,U(1)_c + \frac{3}{2} \,U(1)_d+\frac{3}{2} \,U(1)_e$
\item[$\bullet$] $U(1)_Y= \frac{1}{6}\, U(1)_a + \frac{1}{2} \,U(1)_c + \frac{3}{2} \,U(1)_d+\frac{5}{2} \,U(1)_e$
\item[$\bullet$] $U(1)_Y= \frac{1}{6}\, U(1)_a + \frac{1}{2} \,U(1)_c + x \,U(1)_d+ (x-1)\,U(1)_e$
\item[$\bullet$] $U(1)_Y= -\frac{1}{3}\, U(1)_a - \frac{1}{2} \,U(1)_b  $
\item[$\bullet$] $U(1)_Y= -\frac{1}{3}\, U(1)_a - \frac{1}{2} \,U(1)_b +\frac{1}{2} \, U(1)_e $
\item[$\bullet$] $U(1)_Y= -\frac{1}{3}\, U(1)_a - \frac{1}{2} \,U(1)_b + U(1)_e $
\item[$\bullet$] $U(1)_Y= -\frac{1}{3}\, U(1)_a - \frac{1}{2} \,U(1)_b +\frac{1}{2} \, U(1)_d +\frac{1}{2} \, U(1)_e$
\item[$\bullet$] $U(1)_Y= -\frac{1}{3}\, U(1)_a - \frac{1}{2} \,U(1)_b +\frac{1}{2} \, U(1)_d + U(1)_e$
\item[$\bullet$] $U(1)_Y= -\frac{1}{3}\, U(1)_a - \frac{1}{2} \,U(1)_b +\frac{1}{2} \, U(1)_d +\frac{3}{2} \, U(1)_e$
\item[$\bullet$] $U(1)_Y= -\frac{1}{3}\, U(1)_a - \frac{1}{2} \,U(1)_b + U(1)_d + U(1)_e$
\item[$\bullet$] $U(1)_Y= -\frac{1}{3}\, U(1)_a - \frac{1}{2} \,U(1)_b + U(1)_d + 2\, U(1)_e$
\item[$\bullet$] $U(1)_Y= -\frac{1}{3}\, U(1)_a - \frac{1}{2} \,U(1)_b + x\,U(1)_d + (x-1)\, U(1)_e$
\end{itemize}
Again only a small subset of the hypercharge embeddings exhibit quivers which pass the constarints laid out above and thus give rise to a realistic phenomenology. The quivers are listed in Tables \ref{five stack one}, \ref{five stack two}, and \ref{five stack three}. Below we show that the hypercharge embeddings which contain the free parameter $x$ do not give rise to any solutions.


\begin{table}[h]
\hspace{-.3cm}
\scalebox{.69}{
\begin{tabular}{|c|c|c|c|c|c|c|c|c|c|c|c|c|c|c|c|c|c|}\hline
		\multirow{2}{*}{Solution \#}&\multicolumn{1}{|c}{$q_L$} & \multicolumn{2}{|c}{$d_R$} & \multicolumn{2}{|c}{$u_R$} & \multicolumn{2}{|c}{$L$} & \multicolumn{3}{|c}{$E_R$}
		& \multicolumn{2}{|c}{$H_u$} & \multicolumn{3}{|c|}{$H_d$}\\ \cline{2-16}
		&$(a,\ov{b})$&$(\ov{a},\ov{c})$&$(\ov{a},\ov{d})$
		&$(\ov{a},\ov{e})$&$\Yasymm_a$&$(b,\ov{c})$&$(b,\ov{d})$
		&$(c,e)$&$(\ov{c},e)$&$\ov{\Yasymm}_b$
		&$(\ov{b},\ov{c})$&$(b,e)$
		&$(b,c)$&$(b,\ov{c})$&$(b,d)$
		\\
		\hline\hline			
1&3&0&3&3&0&0&3&0&1&2&0&1&0&1&0\\ \hline
2&3&0&3&3&0&0&3&0&0&3&1&0&1&0&0\\ \hline
3&3&1&2&3&0&2&1&1&0&2&0&1&0&0&1\\ \hline
4&3&0&3&2&1&0&3&0&0&3&1&0&1&0&0\\ \hline
5&3&0&3&1&2&0&3&0&0&3&1&0&1&0&0\\ \hline
6&3&0&3&0&3&0&3&0&1&2&0&1&0&1&0\\ \hline
\end{tabular}}
\caption{\small Spectrum for the quivers with $U(1)_Y= -\frac{1}{3}\, U(1)_a - \frac{1}{2} \,U(1)_b + U(1)_e $.
\label{five stack one}
}
\end{table}
\begin{table}[h]
\hspace{-.25cm}
\scalebox{.73}{
\begin{tabular}{|c|c|c|c|c|c|c|c|c|c|c|c|c|c|c|}\hline
		\multirow{2}{*}{Solution \#}&\multicolumn{1}{|c}{$q_L$} & \multicolumn{1}{|c}{$d_R$} & \multicolumn{3}{|c}{$u_R$} & \multicolumn{1}{|c}{$L$} & \multicolumn{3}{|c}{$E_R$}
		& \multicolumn{2}{|c}{$H_u$} & \multicolumn{3}{|c|}{$H_d$}\\ \cline{2-15}
		&$(a,\ov{b})$&$(\ov{a},\ov{c})$&$(\ov{a},\ov{d})$&$(\ov{a},\ov{e})$&$\Yasymm_a$
	    &$(b,\ov{c})$
		&$(c,d)$&$(c,e)$&${\ov{\Yasymm}}_b$	
		&$(b,d)$&$(b,e)$
		&$(b,c)$&$(\ov{b},\ov{d})$&$(\ov{b},\ov{e})$
		\\
		\hline\hline			
1&3&3&0&3&0&3&0&0&3&1&0&0&1&0\\ \hline
2&3&3&1&2&0&3&0&1&2&0&1&1&0&0\\ \hline
3&3&3&1&2&0&3&1&0&2&1&0&1&0&0\\ \hline
4&3&3&1&2&0&3&0&0&3&0&1&0&0&1\\ \hline
5&3&3&1&2&0&3&0&0&3&1&0&0&1&0\\ \hline
6&3&3&0&2&1&3&0&0&3&1&0&0&1&0\\ \hline
7&3&3&0&1&2&3&0&0&3&1&0&0&1&0\\ \hline
\end{tabular}}
\caption{\small Spectrum for the quivers with $U(1)_Y= -\frac{1}{3}\, U(1)_a - \frac{1}{2} \,U(1)_b + U(1)_d +U(1)_e$.
\label{five stack two}
}
\end{table}
Solutions in Table \ref{five stack three} marked with a  $\diamondsuit$ denote a quivers where an instanton which is required to generate a Yukawa coupling also generates the Weinberg operator. As discussed in section \ref{sec middle scale} in such a scenario the string scale $M_s$ is in the range $10^{9} - 10^{14} \, GeV$. Quivers marked  
with  $\heartsuit$ denote a setups where an instanton which is required to generate a Yukawa coupling also generates the $\mu$-term. As discussed in section \ref{sec lower scale}, in such a case the string scale is further constrained. Setups marked with a $\clubsuit$ denote  quivers where an instanton which generates the Weinberg operator also generates an R-parity violating coupling. For those quivers the string scale should be rather low such that the Weinberg operator inducing instanton suppression is large. Then the simultaneously induced R-parity violating coupling is also highly suppressed and may be compatible with experimental observations. Whether or not the induced R-parity violating coupling indeed poses phenomenological problems needs to be checked carefully.

\begin{table}[h]
\hspace{-1.2cm}
\scalebox{.5}{
\begin{tabular}{|c|c|c|c|c|c|c|c|c|c|c|c|c|c|c|c|c|c|c|c|c|c|c|c|c|c|}\hline
		\multirow{2}{*}{Solution \#}&\multicolumn{2}{|c}{$q_L$} & \multicolumn{3}{|c}{$d_R$} & \multicolumn{3}{|c}{$u_R$} & \multicolumn{4}{|c}{$L$} & \multicolumn{6}{|c}{$E_R$}
		& \multicolumn{6}{|c}{$H_u$} & \multicolumn{1}{|c|}{$H_d$}\\ \cline{2-26}
		&$(a,b)$ & $(a,\ov{b})$ & $(\ov{a},c)$ & $(\ov{a},d)$ & $(\ov{a},e)$ & $(\ov{a},\ov{c})$
		& $(\ov{a},\ov{d})$ & $(\ov{a},\ov{e})$& $(b,\ov{d})$ & $(\ov{b},\ov{d})$ & $(b,\ov{e})$ & $(\ov{b},\ov{e})$
		 & $(c,d) $  & $(c,e) $  & $(d,e) $& $\Ysymm_c$ & $\Ysymm_d$ & $\Ysymm_e$ 
		& $(b,c)$ & $(\ov{b},c)$ & $(b,d)$ & $(\ov{b},d)$& $(b,e)$ & $(\ov{b},e)$&  $(\ov{b},\ov{c})$ \\
		\hline\hline			
	
1&2&1&0&0&3&0&1&2&0&3&0&0&0&0&3&0&0&0&1&0&0&0&0&0&1\\ \hline
2&2&1&0&0&3&0&1&2&0&3&0&0&0&0&1&0&1&1&1&0&0&0&0&0&1\\ \hline
3$^{\heartsuit \clubsuit}$&2&1&3&0&0&1&0&2&1&1&0&1&1&1&0&0&0&1&0&0&0&0&0&1&1\\ \hline
4$^{\heartsuit}$&2&1&0&0&3&1&1&1&0&3&0&0&1&1&0&0&1&0&0&0&0&0&1&0&1\\ \hline
5$^{\heartsuit}$&2&1&0&0&3&1&1&1&0&3&0&0&2&0&1&0&0&0&0&0&0&0&1&0&1\\ \hline
6$^{\heartsuit}$&2&1&0&0&3&1&1&1&0&3&0&0&0&0&1&1&1&0&0&0&0&0&1&0&1\\ \hline
7$^\diamondsuit$&2&1&3&0&0&1&1&1&0&2&1&0&0&2&0&0&1&0&0&0&0&1&0&0&1\\ \hline
8$^\diamondsuit$&2&1&3&0&0&1&1&1&0&2&1&0&2&0&0&0&0&1&0&0&0&1&0&0&1\\ \hline
9$^\diamondsuit$&2&1&3&0&0&1&1&1&0&2&1&0&0&0&0&1&1&1&0&0&0&1&0&0&1\\ \hline
10&2&1&0&3&0&2&0&1&0&0&0&3&0&3&0&0&0&0&1&0&0&0&0&0&1\\ \hline
11&2&1&0&3&0&2&0&1&0&0&0&3&0&1&0&1&0&1&1&0&0&0&0&0&1\\ \hline
12$^{\heartsuit}$&2&1&3&0&0&2&0&1&0&2&1&0&1&2&0&0&0&0&0&1&0&0&0&0&1\\ \hline
13$^{\heartsuit}$&2&1&3&0&0&2&0&1&0&2&1&0&1&0&0&1&0&1&0&1&0&0&0&0&1\\ \hline
14&1&2&0&0&3&0&1&2&3&0&0&0&0&0&3&0&0&0&1&0&0&0&0&0&1\\ \hline
15&1&2&0&0&3&0&1&2&3&0&0&0&0&0&1&0&1&1&1&0&0&0&0&0&1\\ \hline
16$^{\heartsuit}$&1&2&0&0&3&1&0&2&3&0&0&0&0&2&1&0&0&0&0&0&1&0&0&0&1\\ \hline
17$^{\heartsuit}$&1&2&0&0&3&1&0&2&3&0&0&0&0&0&1&1&0&1&0&0&1&0&0&0&1\\ \hline
18&1&2&0&3&0&1&0&2&0&0&3&0&0&2&0&0&0&1&0&0&0&0&1&0&1\\ \hline
19&1&2&0&3&0&1&0&2&0&0&3&0&0&0&0&1&0&2&0&0&0&0&1&0&1\\ \hline
20&1&2&0&3&0&2&0&1&0&0&3&0&0&3&0&0&0&0&1&0&0&0&0&0&1\\ \hline
21&1&2&0&3&0&2&0&1&0&0&3&0&0&1&0&1&0&1&1&0&0&0&0&0&1\\ \hline	
\end{tabular}}
\caption{\small Spectrum for the quivers with $U(1)_Y= \frac{1}{6}\, U(1)_a + \frac{1}{2} \,U(1)_c + \frac{1}{2} \,U(1)_d+\frac{1}{2} \,U(1)_e$.
\label{five stack three}
}
\end{table}

\subsection{Hypercharge: $U(1)_Y = ... + x\, U(1)_d +(x-1) U(1)_e  $
\label{app B}}

In this section we will prove that the two hypercharge embeddings
\begin{align}
\nonumber
U(1)_Y&= \frac{1}{6}\, U(1)_a + \frac{1}{2} \,U(1)_c + x \,U(1)_d+ (x-1)\,U(1)_e \\
\label{eq embeddings parameter}
\\ \nonumber
 U(1)_Y&= -\frac{1}{3}\, U(1)_a - \frac{1}{2} \,U(1)_b + x\,U(1)_d + (x-1)\, U(1)_e
\end{align}
do not give rise to any quivers which satisfy the top-down constraints arising from tadpole cancellation and the presence of the massless $U(1)_Y$. As before, we focus strictly on the case that the quiver exhibits the exact MSSM spectrum. For generic $x$ in both embeddings, the only field charged under the $U(1)_d$ and $U(1)_e$ is the right-handed charged lepton $E_R$, transforming as $(d, \ov e)$. Tadpole cancellation (see eq. \eqref{eq constraint2}) requires that all three families are located at intersections of the stacks $d$ and $e$\footnote{The case that none of the $E_R$'s are charged under the $U(1)_d$ and $U(1)_e$ would also satisfy condition \eqref{eq constraint2}, but then the stacks $d$ and $e$ would not be populated at all, and one would have a three-stack quiver rather than a five-stack quiver.}. Thus, for both embeddings we have all three right-handed charged leptons transforming as $(d, \ov e)$. Now it is easy to show that this in not compatible with the condition \eqref{eq massless constraint abelian}, which ensures that the hypercharge \eqref{eq embeddings parameter} indeed survives as gauge symmetry.

\clearpage \nocite{*}
\bibliography{rev1}
\bibliographystyle{utphys}

\end{document}